\documentclass[journal=jacsat,manuscript=letter]{achemso}

\usepackage[version=3]{mhchem} 

\usepackage{graphicx}
\usepackage{here}
\usepackage{amsmath}

\author{Katharina Boguslawski}
\affiliation[ETH Zurich]
{ETH Zurich, Laboratory of Physical Chemistry, Wolfgang-Pauli-Str. 10, CH-8093 Z\"urich, Switzerland}
\author{Pawe{\l} Tecmer}
\affiliation[ETH Zurich]
{ETH Zurich, Laboratory of Physical Chemistry, Wolfgang-Pauli-Str. 10, CH-8093 Z\"urich, Switzerland}
\author{\"Ors Legeza}
\affiliation["Lend\"ulet" Research Group]
{MTA-WRCP Strongly Correlated Systems "Lend\"ulet" Research Group, H-1525 Budapest, Hungary}
\email{olegeza@szfki.hu}
\author{Markus Reiher}
\affiliation[ETH Zurich]
{ETH Zurich, Laboratory of Physical Chemistry, Wolfgang-Pauli-Str. 10, CH-8093 Z\"urich, Switzerland}
\email{markus.reiher@phys.chem.ethz.ch}

\title[Entanglement Measures for Single- and Multi-Reference Correlation Effects]
{Entanglement Measures for Single- and Multi-Reference Correlation Effects}

\begin{document}
\begin{abstract}
Electron correlation effects are essential for an accurate {\it ab initio} description of molecules. A quantitative \emph{a priori} knowledge
of the single- or multi-reference nature of electronic structures as well as of the dominant contributions to the correlation energy can facilitate the decision
regarding the optimum quantum chemical method of choice. We propose concepts from quantum information theory as
orbital entanglement measures that allow us to evaluate the single- and multi-reference character of any molecular structure in a given orbital
basis set. By studying these measures we can detect possible artifacts of small active spaces.
\end{abstract}

\newpage
The correlation energy is defined as the difference between the ground-state energy of a one-determinant wave function, the Hartree--Fock determinant, and the exact solution
of the Schr\"odinger equation. Qualitatively speaking, it is caused by electronic interactions\cite{Lowdin_corr_energy} beyond the mean-field approach. Even though an \emph{exact} separation of the correlation energy
into individual contributions is not possible, one usually divides correlation effects into three different classes which are denoted \emph{dynamic}, \emph{nondynamic} and
\emph{static}. Although unique definitions of static, nondynamic and dynamic electron correlation do not exist, the dynamic part is considered to be responsible
for keeping electrons apart and is attributed to a large number of configurations, \emph{i.e.}, Slater determinants or configuration state functions, with small (absolute)
coefficients in the wave function expansion, while the nondynamic and static contributions involve only some determinants with large (absolute) weights which are necessary for an appropriate treatment of the quasi-degeneracy of orbitals.\cite{Sinanoglu1963,Bartlett_1994,bartlett_2007}
In particular, static electron correlation embraces a suitable combination of determinants to account for proper spin symmetries and their interactions, whereas nondynamic correlation
is required to allow a molecule to separate correctly into its fragments \cite{Bartlett_1994,bartlett_2007}. 

An accurate treatment of dynamic, nondynamic and static correlation effects is covered by the full configuration interaction (FCI) solution \cite{helgaker2000}. 
However, its steep and unfavorable scaling with the size of the molecule limits the applicability of the FCI approach to systems containing a small number of electrons and small atomic basis sets. 
In order to study larger (chemically interesting) molecules, the FCI wave function needs to be approximated which can be achieved by either single-reference
or multi-reference quantum chemical methods. 

While single-reference approaches like, for instance, M{\o}ller--Plesset perturbation theory or the coupled cluster (CC) ansatz, are able to capture the largest part of the dynamic correlation energy,
the missing nondynamic and static contributions can be recovered by a multi-reference treatment. However, the application of multi-reference methods represents a far more difficult and
computationally expensive task compared to a single-reference study of electronic structures.
Furthermore, employing any wave-function based electron correlation approach requires some \textit{a priori} knowledge about the interplay
of dynamic, nondynamic and static electron correlation effects. 

Over the past few decades, a number of different diagnostic tools have been developed to characterize the single- or multi-refence nature of molecular systems in order to validate the quality
and performance of single-reference quantum chemical methods. 
For instance, if the absolute or squared weight of the reference configuration (the $C_0$ coefficient) obtained from a CI calculation are above a certain threshold ($C_0>0.95$
or $C_0^2>0.90$), the electronic structure is considered to be of single-reference nature \cite{lee_1989}.
As an alternative measure, Lee \textit{at al.} \cite{Lee_t1_1,lee_1989,T1_open-shell} proposed to analyze the Euclidean norm of the $t_1$ amplitudes optimized in a CC calculation
which is usually denoted as $T_1$ diagnostics. 
It was shown that single-reference CC can be considered accurate when the $T_1$ diagnostic is smaller than $0.02$ for main group elements \cite{Lee_t1_1,lee_1989,Jannsen_T1} and 0.05
for transition metals \cite{Sears2008,Sears2008a,MR_test_Wilson} and actinide compounds \cite{real09,pawel1}, respectively. 
Since, however, the above mentioned criteria have not been rigorously defined and turn out to be system- and method-dependent, additional measures abbreviated by $D_1$ and $D_2$ which are
based on single and double excitations were introduced to assess the quality of a single-reference CC calculation \cite{Nielsen1999,Leininger2000,Lee_T1_and_D1}.
Although such diagnostic tools can reveal deeper insights into the electronic structure of molecules, they present an \emph{a posteriori} analysis of the performance of quantum chemical calculations.
So far, no universal measures have been introduced to quantify the electron correlation effects in a most general fashion that can be applied to both single-reference and multi-reference problems.

Since electron correlation effects are caused by the interaction of electrons that occupy specific orbitals used to construct the Slater determinant basis, an intuitive way
to study electron correlation is to measure the interaction among any pair of orbitals or the interaction of one orbital with the remaining ones which are incorporated in a FCI wave function.
Thus, a universal procedure of quantifying the entanglement between orbitals is sought under the constraint that no artificial truncation of the complete $N$-particle Hilbert space is
performed. The latter is required to exclude any method-dependent error in the diagnostic analysis like the restriction to a predefined excitation hierarchy or to some zeroth-order wave function.  
An efficient approach to systematically approximate the FCI solution even for large molecules and complicated electronic structures is to apply conceptually different electron correlation methods
like the density matrix renormalization group (DMRG) algorithm developed by White \cite{white}. DMRG allows us to treat large active orbital spaces \cite{marti2010b,chanreview} without a
predefined truncation of the complete $N$-particle Hilbert space.

The interaction between orbitals can then be calculated employing concepts from quantum information theory. Different entanglement measures have been introduced into the DMRG algorithm
almost a decade ago which are routinely applied in order
to accelerate DMRG convergence towards the global energy minimum and paved the way for black-box DMRG calculations \cite{orbitalordering}. In this respect, some of us 
\cite{legeza_dbss} employed the von Neumann entropy for subsystems containing one single orbital to quantify the correlation between this orbital and the remaining set of orbitals contained in the active space.
The single orbital entropy $s(1)_i$ can be determined from the eigenvalues of the reduced density matrix $\omega_{\alpha,i}$ of a given orbital $i$,
\begin{equation}\label{Eq:single-orbital}
s(1)_i = - \sum_\alpha \omega_{\alpha,i} \ln \omega_{\alpha,i},
\end{equation}
while the total quantum information encoded in the wave function \cite{legeza_dbss3} reads
\begin{equation}\label{Eq:single-orbital}
I_{\rm tot} = \sum_i s(1)_i.
\end{equation}
In order to allow a balanced treatment of electron interaction, 
Rissler \emph{et al.} \cite{Rissler2006519}
presented a scheme to determine the informational content of any pair of orbitals using the von Neumann entropy. The so-called mutual information $I_{i,j}$ quantifies the
correlation of two orbitals embedded in the environment comprising all other active orbitals,
\begin{equation}\label{Eq:mutual-info}
I_{i,j} = s(2)_{i,j} - s(1)_{i} - s(1)_{j},
\end{equation}
where $i = {1\ldots k}$ is the orbital index and runs over all $k$ one-particle states, $\omega_{\alpha,i}$ is the $\alpha$ eigenvalue of the reduced density matrix
of orbital $i$ \cite{legeza_dbss}, while $s(2)_{i,j}$ is the two-orbital entropy between a pair ${i,j}$ of sites \cite{Rissler2006519}.
Thus, the single orbital entropy and the mutual information represent convenient measures of entanglement and due to their general definition they can be employed to
quantify different types of correlation present in arbitrary quantum chemical systems.   

\begin{figure}[h]
\centering
\includegraphics[width=0.6\linewidth]{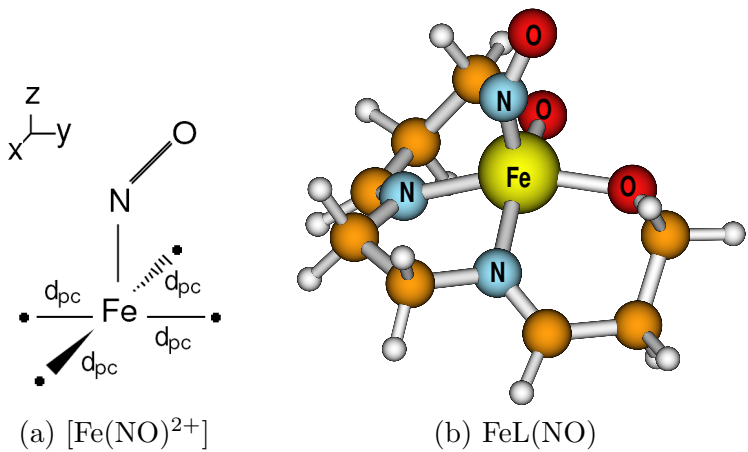}
\caption{Structures of the bare and ligated iron nitrosyl complexes, [Fe(NO)]$^{2+}$ and FeL(NO), respectively.
The [Fe(NO)]$^{2+}$ molecule is surrounded by four point charges of $-0.5$ e at a distance of $d_{\rm pc} = 1.131$ \AA{} from the iron center. For FeL(NO), a structure optimization
was performed enforcing $C_{2v}$ symmetry (BP86/TZP).
}\label{fig:structures}
\end{figure}

\begin{figure}[h]
\centering
\includegraphics[width=0.9\linewidth]{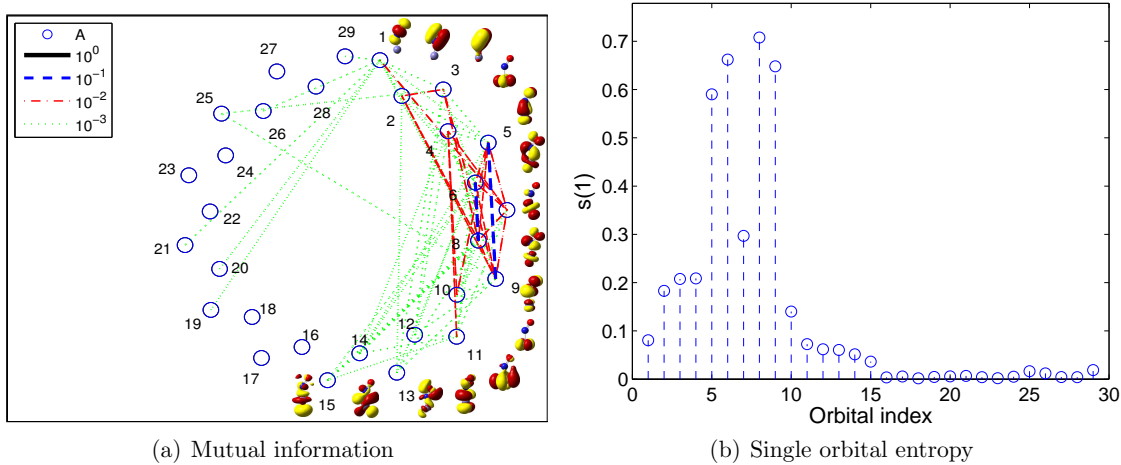}
\caption{Mutual information and single orbital entropies $s(1)$ for a DMRG(13,29) calculation employing the DBSS approach with a minimum and maximum number of renormalized active system states
set to 128 and 1024, respectively, and a quantum information loss of $10^{-5}$ for the [Fe(NO)]$^{2+}$ molecule surrounded by four point charges at a distance of $d_{\rm pc} = 1.131$ \AA{} from the iron center.
The orbitals are numbered and sorted according to their (CASSCF) natural occupation numbers. Each orbital index in (b) indicates one molecular orbital.
The orbital index in (b) and the orbital number in (a) correspond to the same natural orbital. The total quantum information is $I_{\rm tot}=4.103$.
}\label{fig:feno-info}
\end{figure}

To demonstrate our approach, we consider the [Fe(NO)]$^{2+}$ molecule embedded in a point charge field (see Fig.\ \ref{fig:structures}(a)) which we recently identified
as a difficult system for standard electron correlation approaches since it requires a balanced treatment of both static, nondynamic and dynamic correlation effects
\cite{feno,fenoDMRG}. \ref{fig:feno-info} shows the mutual
information and single orbital entropies obtained from a DMRG calculation where 13 electrons have been correlated in 29 orbitals (see also Ref.\ \citenum{fenoDMRG} for further details).
Those active orbitals which are important for static and nondynamic electron correlation are displayed in the Figure together with four double-$d$-shell orbitals.
By examining \ref{fig:feno-info}, we can distinguish three different interaction strengths of the mutual information which are indicated by blue, red and green lines.
The blue lines connect each bonding and antibonding combination of the Fe 3$d$- and NO $\pi^*$-orbitals (there are two of them), while the red lines connect
orbitals which are usually included in standard electron correlation calculations to account for static correlation effects.

A similar pattern can be observed for the single orbital entropies where we can distinguish three different orbital blocks: orbitals (i) with large single orbital
entropies ($>0.5$), (ii) with medium-sized single orbital entropies ($0.1< s(1)_i < 0.5$) and (iii) with small or almost zero single orbital entropies ($0< s(1)_i  <0.1$).
Furthermore, orbitals which are strongly entangled with at least one other orbital (blue and red lines in the mutual information diagram of \ref{fig:feno-info}) correspond to large single orbital
entropies. Thus, we can identify orbitals with \emph{both} large $I_{i,j}$ and large $s(1)_i$ to be important for nondynamic correlation effects, while intermediate values
of $I_{i,j}$ and $s(1)_i$ indicate orbitals which are crucial for a correct description of static electron correlation. The (dominant) part of the dynamic
electron correlation energy is then captured by excitations into orbitals which are weakly entangled with all other orbitals (green lines in the mutual information diagram of \ref{fig:feno-info}
and small $s(1)_i$). Note that the double-shell orbitals indeed recover a large part of the dynamic correlation energy (largest values for $s(1)_i$ in the third block).
However, it was shown by us that the missing dynamic correlation effects induced by the remaining virtual orbitals are of considerable importance for an accurate
description of the electronic structure of the [Fe(NO)]$^{2+}$ molecule \cite{fenoDMRG}. Since these orbitals are solely connected by green lines and correspond to small $s(1)_i $, their contribution
to the electronic energy is of pure dynamic nature and thus a standard CASSCF model of the electronic structure of [Fe(NO)]$^{2+}$ is not sufficient. Note that this could be cured
by, for instance, applying second-order-perturbation theory upon a CASSCF reference function \cite{caspt21,caspt22}.

\begin{figure}[h]
\centering
\includegraphics[width=0.9\linewidth]{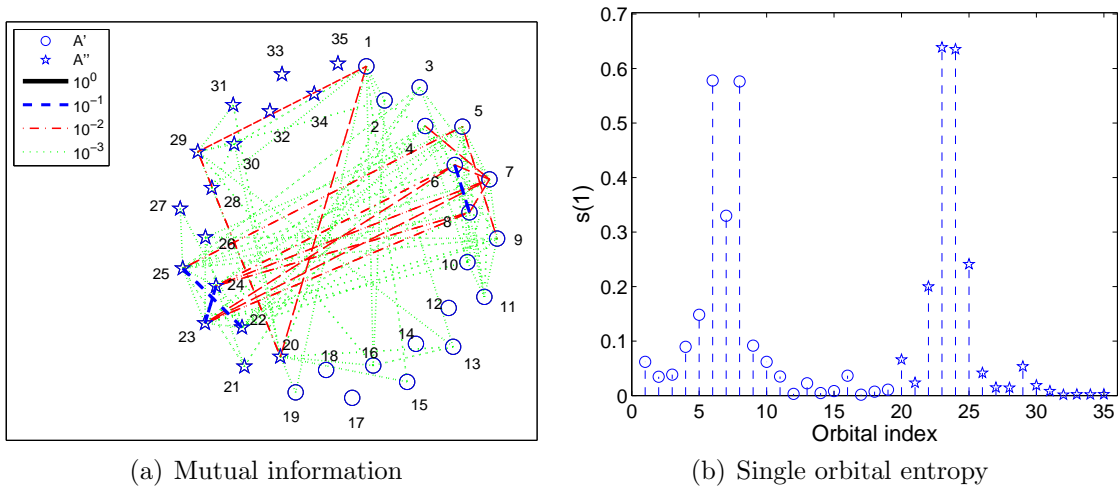}
\caption{Mutual information and single orbital entropies $s(1)$ for a DMRG(21,35) calculation employing the DBSS approach with a minimum and maximum number of
renormalized active system states of 128 and 1024, respectively, and a quantum information loss of $10^{-5}$ for the ligated iron nitrosyl complex as displayed in \ref{fig:structures}(b).
The orbitals are numbered and sorted according to their (CASSCF) natural occupation numbers and sorted according to their irreducible representation. Each orbital index in (b)
indicates one molecular orbital.
The orbital index in (b) and the orbital number in (a) correspond to the same natural orbital.
}\label{fig:ligated-model}
\end{figure}

Next, we can investigate the changes in the electronic structure when the point charge environment is substituted by one or several ligand molecules. We replace
the point charges by a small model molecule of a salen ligand where the aromatic rings have been substituted by CH$_2$ units as displayed in \ref{fig:structures}(b).
The mutual information and single orbital entropies obtained from a DMRG calculation correlating 21 electrons in 35 molecular orbitals and imposing $C_{2v}$ symmetry are shown in \ref{fig:ligated-model}.
In particular, we obtain similar entanglement diagrams for the ligated iron nitrosyl compound as found for the small
[Fe(NO)]$^{2+}$ molecule, \emph{i.e.}, three groups of orbitals which can be classified by their (combined) $I_{i,j}$ and $s(1)_i$ contributions. Again, the two bonding
and antibonding combinations of the Fe 3$d$- and NO $\pi^*$-orbitals (\#23--\#24 and \#6--\#8) are important for a correct description of nondynamic correlation effects.
Yet, in contrast to the bare [Fe(NO)]$^{2+}$ complex, the Fe 3$d_{xy}$-orbital now interacts with one ligand $\sigma$-orbital and its bonding and antibonding combinations are
strongly entangled (\#22--\#25). The single orbital entropy profile further indicates that these orbitals need to be considered for an accurate treatment of static correlation
and are thus, together with the remaining highly entangled orbitals (\#4 to \#9), \emph{i.e.}, those that would have been included in any standard (minimum) active space calculation,
mandatory to capture the static correlation energy.
However, we now observe a great number of orbitals which are weakly entangled and which comprise (very) small single orbital entropies. Hence, the influence of dynamic
correlation increases after ligation and can be referred to the additional virtual ligand orbitals that are available for possible excitations.
Simultaneously, the contribution of static electron correlation decreases which can be explained by the reduced number of statically entangled orbitals (the red lines in
\ref{fig:ligated-model}) and the smaller single orbital entropies (compare \ref{fig:ligated-model} and \ref{fig:feno-info}).
Under the process of ligation, the multi-reference character of the electronic wave function depletes and thus we must be able to properly account for dynamic
correlation effects in a quantum chemical description of the ligated iron nitrosyl compound.

\begin{figure}[H]
\centering
\includegraphics[width=0.8\linewidth]{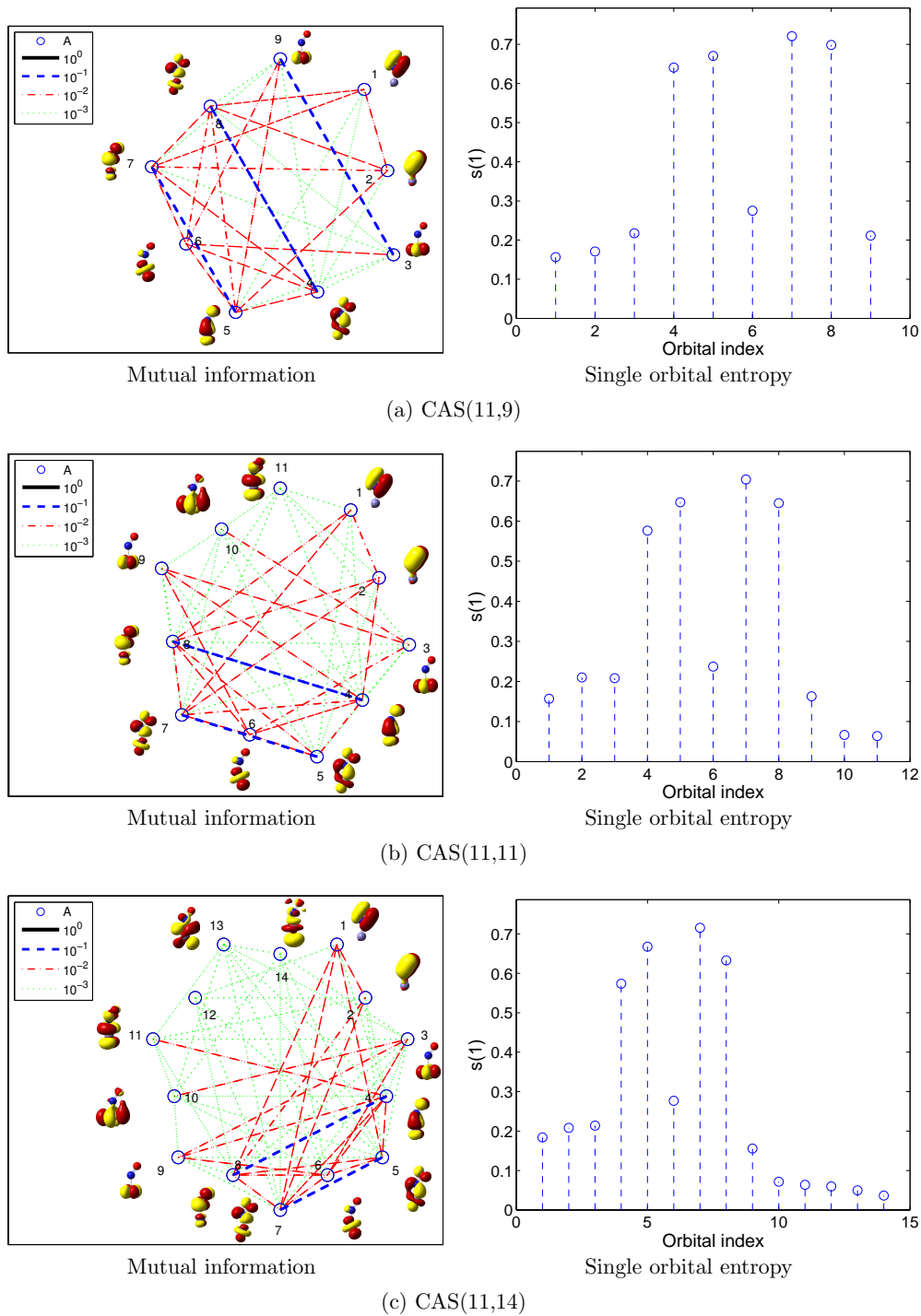}
\caption{Mutual information and single orbital entropies $s(1)$ for DMRG(11,$y$) calculations determined for different
numbers of active orbitals in [Fe(NO)]$^{2+}$ surrounded by four point charges at a distance of
$d_{\rm pc} = 1.131$ \AA{} from the iron center. For each DMRG calculation the number of renormalized active-system states was increased until the CASSCF reference energy was obtained.
The orbitals are numbered and sorted according to their (CASSCF) natural occupation numbers. Each orbital index
indicates one molecular orbital. 
The orbital index in (b) and the orbital number in (a) correspond to the same natural orbital. For CAS(11,9), CAS(11,11) and CAS(11,14), the total quantum information
corresponds to 3.761, 3.678 and 3.909, respectively.
}\label{fig:small-cas}
\end{figure}

The single orbital entropies and mutual information patterns can be employed to analyze possible artifacts which emerge from space calculations with active spaces that are too small.
\ref{fig:small-cas} summarizes the $I_{i,j}$ and $s(1)_i$ diagrams obtained for different choices of the active space of the [Fe(NO)]$^{2+}$ molecule embedded
in a point charge field. For the CAS(11,9) calculation, static and nondynamic electron correlation effects are overestimated compared to the DMRG(13,29)
reference calculation (note the increased number of blue and red lines as well as the larger values for $s(1)_i$). Simultaneously, the contributions from
dynamic correlation are decreased. These artifacts can be---at least to some extent---resolved if the active space is enlarged. Including additional
Fe $3d$-orbitals into the active space in a CAS(11,11) calculation partially corrects the description of static correlation (\emph{cf.} orbitals \#3 and \#9
are less entangled), but does not account for an accurate description of dynamic correlation, \emph{e.g.}, orbitals \#6 and \#9 still appear important for static electron
correlation. In the DMRG(13,29) reference calculation, however, they are connected solely through dynamic correlation effects, and thus the static correlation energy
is overemphasized. If two additional double-shell orbitals are included upon the CAS(11,11) calculation, a great part of the dynamic correlation energy can be accounted for
in a standard CAS(11,14)SCF calculation (note that one additional virtual ligand orbital was rotated in the active space despite the four Fe $3d$-double shell orbitals).
Nevertheless, the pattern in the mutual information does not improve compared to the CAS(11,11) calculation and the static correlation energy is still overestimated
(\emph{cf.} orbitals \#6 and \#9 remain strongly entangled).

Similar conclusions can be drawn when analyzing the evolution of the single orbital entropies with respect to the dimension of the active orbital space. For CAS(11,9),
the $s(1)_i$ values are in general too large which coincides with the overestimation of static and nondynamic correlation effects, while the dynamic contributions
are diminished. These artifacts can be corrected by extending the dimension of the active orbital space and thus allowing for a better treatment of dynamic
correlation. However, if the nondynamic correlation energy is overestimated, the static electron correlation contributions will be underestimated,
which can be observed in too small single orbital entropies compared to a FCI reference, and \emph{vice versa} which results in too large values for $s(1)_i$.
Although the wrong estimate of static and nondynamic electron correlation caused by a too small dimension of the active orbital space can be corrected by including
additional virtual orbitals like Fe 3$d$-double-shell orbitals, the improvements are insufficient since the contributions of these virtual orbitals
to the dynamic correlation energy remain underestimated compared to the DMRG(13,29) reference calculation.
Furthermore, extending the active space from CAS(11,11) over CAS(11,14) to CAS(13,29)
leads to $I_{\rm tot}$ values of 3.678, 3.909 and 4.103, respectively. Hence, more (dynamic) electron correlation is recovered when the active space is increased.
For CAS(11,9), however, dynamic correlation is considerably underestimated which leads to overrated static correlation effects and results in a too large value for the total quantum information
of $I_{\rm tot}=3.761$ when compared to the CAS(11,11) calculation.

In this work, we have presented a quantitative measure to assess electron correlation effects which are independent of the reference wave function and do not require
an \emph{a priori} knowledge about the single- or multi-refence character of the electronic structure. In our analysis, the DMRG algorithm was employed which allows us
to systematically approach the FCI solution. The static, nondynamic and dynamic contributions to the
correlation energy can be distinguished by examining the entanglement patterns of orbitals. We demonstrated that the single- or multi-refence nature of electronic structures
is encoded in the mutual information and single orbital entropies. These quantities do not significantly depend upon the accuracy
of our DMRG calculations and can be already obtained from fast and inexpensive DMRG sweeps. The cost for these DMRG sweeps needed to acquire the
entanglement measures is thus negligible. Expensive in terms of computing time is the calculation of the two-electron integrals in the molecular orbital basis,
which, however, is a mandatory step in any correlation treatment and thus a prerequisite of the correlation treatment chosen after the evaluation of the entanglement measures.
Of course, the DMRG sweeping may also be continued until convergence is reached if an alternative like a CC model is not expected to yield more accurate results or in a shorter time, respectively.
The entanglement analysis proposed here can be performed in any orbital basis without loss of generality and can provide insights which quantum chemical method to choose for an accurate description of
the molecule under study. Furthermore, we highlighted the artifacts emerging from small active space calculations.

\section{Computational Details}
All DMRG calculations as well as the calculation of the mutual information and single orbital entropies have been performed with the Budapest DMRG program 
\cite{dmrg_ors}. As orbital basis, the natural orbitals obtained from preceding CASSCF calculations employing the \textsc{Molpro} program package \cite{molpro} in a cc-pVTZ basis
set \cite{dunning,dunning2} are used comprising 11 electrons correlated in 14 orbitals (for both the bare and ligated iron nitrosyl complex). In order to accelerate convergence,
the dynamic block state selection (DBSS) approach \cite{legeza_dbss2,legeza_dbss3} and the dynamically extended active space procedure \cite{legeza_dbss} were employed, while the orbital ordering
was optimized for each active space calculation according to Ref.\ \citenum{orbitalordering}.
The small active space calculations, {i.e.}, CAS(11,9), CAS(11,11) and CAS(11,14), are performed in the corresponding CASSCF natural orbital basis.

\acknowledgement

The authors gratefully acknowledge financial support by the Swiss national science foundation SNF (project 200020-132542/1),
and from the Hungarian Research Fund (OTKA) under Grant No.~K73455 and K100908.
K.B.\ thanks the Fonds der Chemischen Industrie for a Chemiefonds scholarship.
{\"O}.L.\ acknowledges support from the Alexander von Humboldt foundation and from ETH Zurich
during his time as a visiting professor.


\providecommand*\mcitethebibliography{\thebibliography}
\csname @ifundefined\endcsname{endmcitethebibliography}
  {\let\endmcitethebibliography\endthebibliography}{}

\end{document}